\documentclass[preprint,review,10pt]{elsarticle}

\usepackage[margin=3cm]{geometry}
\usepackage[noend]{algpseudocode}
\usepackage[english]{babel}
\usepackage{verbatim}
\usepackage[table]{xcolor}
\usepackage{graphicx}
\usepackage{svg}
\usepackage{caption}
\usepackage{subcaption}
\usepackage{amssymb}
\usepackage{amsmath}
\usepackage{amsfonts}
\usepackage{mathrsfs}
\usepackage{float}
\usepackage{multirow}
\usepackage{tabularx}
\usepackage{array}
\usepackage{color}
\usepackage{leftidx}
\biboptions{numbers,sort&compress}
\usepackage{xcolor}
\usepackage{soul}
\usepackage[]{algorithm2e}
\usepackage{makecell} 
\usepackage{url}
\usepackage{todonotes}
\definecolor{MC}{rgb}{0.0, 0.5, 1.0}

\usepackage[colorlinks,citecolor=red,urlcolor=blue,bookmarks=false,hypertexnames=true]{hyperref} 

\newcommand{\eg}{\textit{e}.\textit{g}. }

\newcommand*{\colorboxed}{}
\def\colorboxed#1#{%
  \colorboxedAux{#1}%
}
\newcommand*{\colorboxedAux}[3]{%
  \begingroup
    \colorlet{cb@saved}{.}%
    \color#1{#2}%
    \boxed{%
      \color{cb@saved}%
      #3%
    }%
  \endgroup
}

\newcolumntype{C}{ >{\centering\arraybackslash} m{2cm} }
\newcolumntype{F}{ >{\centering\arraybackslash} m{2.5cm} }
\newcolumntype{D}{ >{\centering\arraybackslash} m{1.5cm} }

\usepackage{dashbox}

\usepackage{tikz}

\usepackage{rotating}

\usepackage{lineno}

\newcommand{\RN}[1]{%
  \textup{\uppercase\expandafter{\romannumeral#1}}%
}

\definecolor{azure(colorwheel)}{rgb}{0.0, 0.5, 1.0}

\definecolor{mygreen}{rgb}{0.0, 0.5, 0.0}
\definecolor{darkraspberry}{rgb}{0.53, 0.15, 0.34}
\definecolor{bleudefrance}{rgb}{0.19, 0.55, 0.91}

\newcolumntype{B}[1]{>{\centering\arraybackslash}m{#1}}

\begin{document}

\begin{frontmatter}


\title{Recurrent neural networks and transfer learning for elasto-plasticity in woven composites}

\author[mymainaddress]{Ehsan Ghane}
\author[mythirdaddress]{Martin Fagerström}
\author[mymainaddress]{Mohsen Mirkhalaf\corref{mycorrespondingauthor}}

%
\cortext[mycorrespondingauthor]{Email: mohsen.mirkhalaf@physics.gu.se}
\address[mymainaddress]{Department of Physics, University of Gothenburg, Gothenburg, Sweden}
\address[mythirdaddress]{Department of Industrial and Materials Science, Chalmers University of
Technology, Gothenburg, Sweden}

\begin{abstract}
\noindent Woven composites exhibit complex meso-scale behavior depending on meso- and micro-structural parameters. Accurately modeling their mechanical response is challenging and computationally demanding, especially for inelastic behavior. To address the computational burden, we have developed a Recurrent Neural Network (RNN) model as a surrogate for meso-scale simulations. As a basis for RNN training, a mean-field model generates a comprehensive data set representing elasto-plastic behavior. Arbitrary six-dimensional time histories of strain are used to generate multiaxial stress-strain histories under random walking and cyclic loading conditions as the source and target tasks, respectively. First, the RNN model is trained for the source task. The same model is trained leveraging \textit{transfer learning} for the target task, containing fewer data and sparse features because only some strain components are non-zero. The candidate model is successfully trained and validated through a grid search exploration of over 220 different RNN configurations and demonstrates accurate predictions for both source and target tasks. The results demonstrate that transfer learning could be used to train the RNN effectively under varying strain conditions and arbitrary constituents' material properties, suggesting its potential as an appropriate tool for modeling path-dependent responses in woven composites. 
\end{abstract}


\end{frontmatter}
\section{Introduction}
\label{Introduction}
Woven composite laminates are commonly used in structural applications due to their automated and cost-effective manufacturing processes. 
However, modeling of woven composites presents significant challenges due to the presence of two heterogeneous sub-scales, the meso-scale, and the micro-scale, and the intricate interlacing of yarns, resulting in the development of complex stress states \cite{doitrand2017mesoscale}. 

In order to predict the complex behavior of woven composites governed by the heterogeneous sub-scales configuration, different full-field micro-mechanical and meso-scale models have been developed (\eg \cite{ma2021analysis, doitrand2015comparison, doitrand2017mesoscale}). However, one of the major challenges
of using meso-scale models is their high computational cost, which hinders the usage of these models for engineering applications \cite{spilker2023three}. As a remedy, mean-field models have been proposed and used (\eg \cite{wu2021per}). In these models, average stress and average strain are considered for each sub-scale constituent. A better computational performance (compared to full-field models) is obtained, although at the expense of lower fidelity and accuracy. 

Recently, data-driven approaches have gained considerable interest in developing surrogate models for different composites (\eg \cite{rocha2021fly, Mentges2021, maia2023physically, el2023predicting, bessa_framework_2017, dekhovich2023cooperative}). Different kinds of Artificial Neural Networks (ANNs) have been used to develop remarkably efficient and highly accurate surrogate models. A feed-forward architecture is typically good enough to develop an ANN-enhanced model in the linear elastic regime \cite{ghane2023multiscale}. However, for inelastic path-dependent behavior, it is required to use more advanced ANN architectures \cite{rosenkranz2023comparative}. In recent years, different kinds of Recurrent Neural Networks (RNNs),  such as Gated Recurrent Units (GRU)\cite{cho2014learning} and Long Short-Term Memory (LSTM) networks \cite{hochreiter1997long}, have been employed for the inelastic path-dependent behavior of different composite materials (see \eg \cite{Mozaffar2019,WU2022,friemann2023micromechanics,maia2023physically}). A remarkable computational enhancement and a high level of accuracy were obtained.

Training RNNs is sensitive to initializing the training parameters (weights and thresholds) due to the risk of vanishing or exploding gradients \cite{glorot2010understanding}. Vanishing gradients result in slow learning while exploding gradients lead to unstable training. 
Some methods have been suggested for initializing layer weights and biases \cite{he2015delving, glorot2010understanding}. These methods can greatly impact how well the deep network trains. However, they are still based on randomly initializing the training parameters. 

Furthermore, RNNs encounter challenges when dealing with sparse feature data sets where inputs have many zero values. For instance, in the case of a stress prediction task using strain tensor components as inputs, when we have pure-shear loading, only one component of the input feature is non-zero. This results in data sparsity that can interfere with learning and hinder the network's ability to capture meaningful patterns. 

This paper employs a physics-guided initialization \cite{benady2023nn} of weights and thresholds using \textit{transfer learning} \cite{yang2020transfer}. Using transfer learning, RNNs are able to overcome initialization challenges by leveraging knowledge derived from previously trained models. 
The network is initialized with the pre-trained model containing the expected material parameters, known as the source task. The network is then fine-tuned in accordance with the target task. This study's source task network is to predict the full stress history of nonlinear woven composites with different constituent properties subjected to random multiaxial strain histories. The target task is to predict stress histories under conventional cyclic loading. Models that are trained on dense or diverse data sets can provide more robust representations of features that can be generalized well to sparse data sets. This approach accelerates training, enhances generalization over sparse feature samples, and facilitates effective learning.

This study formulates various GRU and LSTM models to predict the elasto-plastic behavior of woven composites. The specific nature of the task determines the choice between GRU and LSTM architectures. Moreover, the optimal performance of these models is closely linked to the number of training parameters. This facet is evaluated in depth within the framework of this study. 

To generate the two required data sets, a mean-field model from Digimat-MF \cite{Digimat-MF} is employed. The matrix and reinforcements are considered elastic-plastic and elastic, respectively. Notably deviating from prevailing trends observed in developing material-specific ANN surrogate models for composites (\eg \cite{dornheim2023neural, maia2023physically}), this study uniquely incorporates many matrix and reinforcement properties into its framework. 

Six-dimensional arbitrary loading paths (for six independent strain components) are generated and applied to the meso-structural simulations. The simulation results serve as a data set for the source task. Bi-axial and pure shear cyclic load paths are generated for the target task data set, each containing a different peak strain, strain ratio, load ratio, and number of cycles. 
As a result, two comprehensive data sets (including stress-strain responses) for generic woven composites (varying matrix and reinforcement properties) subjected to randomly sampled and cyclic loading histories are generated. 

When training a neural network, highly sensitive parameters and hyperparameters are present, such as learning rate, minibatch size, regularization strength, dropout rate, and network architecture \cite{pascanu2013difficulty}. The identification of the best combination of hyperparameters requires a comprehensive study, which is frequently neglected throughout the literature due to the time-consuming nature of the process \cite{Mozaffar2019, abueidda2021deep, huang2020machine, vlassis2023geometric, jones2022neural, friemann2023micromechanics}. 
We attempt to determine an optimal network by testing many possible combinations. The results show that an LSTM model is successfully trained and validated on the target task, enabling highly efficient elasto-plastic path-dependent simulations. 

The remainder of this paper is structured as follows. Section \ref{dataGeneration} describes the data generation process, including the sub-scale mean-field modeling approach, material constituents, and design of computational experiments. Section \ref{RNN} details the RNN model design and training and transfer-learning strategy. The obtained results and comparisons to micro-mechanical simulations are presented in Section \ref{Results and discussion}, followed by a discussion of the developed RNN model. Concluding remarks are provided in Section \ref{Conclusions}. 

\section{Data generation}
\label{dataGeneration}

Every data sample contains (i) a particular set of constituent material properties, (ii) a 6D random strain loading path, and (iii) a 6D time history of homogenized stress components. The database is generated, considering various material parameters and loading conditions, to capture the complex behavior of various woven composites subjected to complex strain states. Two comprehensive data sets (one for the random walk \footnote[1]{It's worth noting that the term "random walk" is often associated with stochastic processes, where an unpredictable element determine the next step or state. Random walks in strain loading suggest that a strain value evolves over time as a result of random factors, resulting in a pattern that may resemble a walker's path.} loading and one for the cyclic loads) are created by carefully controlling the input variables and using mean-field simulations as described below. 
\subsection{Meso-scale simulations}
\label{mesocsaleModelingApproach}
A mean-field model which uses the Mori-Tanaka theory for homogenization, implemented in \textsc{Digimat-MF}, is used to conduct non-linear path-dependent elasto-plastic simulations of woven composites with varying properties. While the geometry of the meso-scale structure remains unchanged, the micro-structural constituent's properties vary in each virtual sample (see more details in Subsection \ref{design of experiments} below).
As a consequence, the mean-field homogenization procedure involves two steps. In the first step, the sub-scale of the composite material being studied is divided into smaller units called pseudo-grains (PGs). Each PG represents a localized region within the composite. 
Once the division into PGs is completed, the homogenization process begins. Each pseudo-grain is individually subjected to a homogenization procedure, where the behavior and properties of the constituent materials within the PG are analyzed. 
In the second step, homogenization is extended to the entire sub-scale. The effective response of the entire sub-scale is computed by averaging over the collective behavior of all the homogenized pseudo-grains. 
An interested reader is referred to \cite{Doghri2005} for a more comprehensive understanding of the modeling approach.

\subsubsection{Constitutive behavior of sub-scale phases }
\label{MatrixPhaseConstitutive law}
Polymeric materials typically show a strain rate-dependent mechanical response (see, \eg \cite{mirkhalaf2016elasto,mirkhalaf2017modelling}). However, an approximation of rate-independent behavior could be considered for most thermoset polymeric materials under quasi-static loading rates and at room temperature. Therefore, this study considers a rate-independent elasto-plastic response for the matrix material.

The matrix is assumed to obey $J_2$-plasticity with linear-exponential hardening \cite{simo2006computational}. 
The yield function is given by
\begin{equation}
    \label{eq:yieldfunction}
    \Phi({\sigma},\kappa) = \sigma_{\text{eq}}-(\sigma_{\text{y}}+\kappa)\leq 0,
\end{equation}
where $\sigma_{\text{y}}$ is the yield stress, and $\sigma_{\text{eq}}$ is the von Mises equivalent stress defined by
\begin{equation}
    \label{eq:vonMisesSigma}
    \sigma_{\text{eq}} = \sqrt{\frac{3}{2}{\boldsymbol{\sigma}}_{\text{dev}}:{\boldsymbol{\sigma}}_{\text{dev}}},\quad {\boldsymbol{\sigma}}_{\text{dev}} = \boldsymbol{{\sigma}}-\frac{1}{3}\text{tr}({\boldsymbol{\sigma}})\mathbf{I}.
\end{equation}
In Equation (\ref{eq:vonMisesSigma}),  ${\boldsymbol{\sigma}}_{\text{dev}}$ is the deviatoric stress tensor, and $\mathbf{I}$ is the second-order identity tensor. In Equation (\ref{eq:yieldfunction}), $\kappa$ is the hardening stress which is given by
\begin{equation}
\label{eq:microhardeningstress}
    \kappa= H \bar{\varepsilon^{\text{p}}} +H_{\infty}\left(1-e^{-m\bar{\varepsilon^{p}}}\right),
\end{equation}
where $H$ is the linear hardening modulus, $H_{\infty}$ is referred to the hardening modulus, $m>0$ is the hardening exponent, and $\bar{\varepsilon}^{\text{p}} \geq 0$ is the accumulated plastic strain.

Reinforcements are assumed to be isotropic and linearly elastic and obey Hooke's generalized law. Furthermore, the matrix and reinforcement phases are assumed to be perfectly bonded. 
Despite the fact that this assumption may not be true in all cases, it provides a reasonable basis for examining the overall behavior of the woven composite under study.

\subsection{Design of computational experiments}
\label{design of experiments}

This study involves two sets of input features for computational experiments: (i) static features representing fiber and matrix material properties and (ii) multi-dimensional sequential load path components (Subsection \ref{loadingPathGenerator}).
A wide range of properties are considered for the static features, which are given in Table \ref{tab:materialparameters}.
\begin{table}[h!]
    \centering
    \caption{Ranges of material parameters used to generate the required data set during simulations. All data samples in the training and testing sets are distinguished by a unique set of properties of their micro-structural constituents.} 
    \scalebox{0.85}{
        \begin{tabular}{| c | c c |}
            \hline
            & \textbf{Parameter} & \textbf{Value}\\ \hline \hline
            \multirow{3}{*}{\textbf{Fiber}} & Young's modulus $E_{\text{F}}$ & 69-700 GPa\\ \cline{2-3}
            & Poisson's ratio $\nu_{\text{F}}$ & 0.25-0.49\\ \cline{2-3}
            & Fiber volume fraction $V_f$ & $0.10-0.48$\\ \cline{2-3} \hline \hline
            \multirow{6}{*}{\textbf{Matrix}} & Young's modulus $E_{\text{M}}$ & 2-10 GPa\\ \cline{2-3}
            & Poisson's ratio $\nu_{\text{M}}$ & 0.2-0.49\\ \cline{2-3}
            & Yield stress $\sigma_{\text{y}}$ & 31-66 MPa\\ \cline{2-3}
            & Linear hardening modulus $H$ & 1-200 MPa\\ \cline{2-3}
            & Hardening modulus $H_{\infty}$ & 10-30 MPa\\ \cline{2-3}
            & Hardening exponent $m$ & 1-500\\ \hline
        \end{tabular}
    }
    \label{tab:materialparameters}
\end{table}

\subsubsection{Sampling material features}
\label{SamplingMaterialFeatures}
Having uniformly distributed input features to train an ANN properly is beneficial. Regular grids of sample points can lead to coincident projections in different hyper-planes  \cite{bessa_framework_2017}, negatively impacting machine learning, especially in high-dimensional spaces \cite{bishop2006pattern}. Thus, using an effective sampling technique helps to achieve a random and uniform distribution while reducing simulation costs.

Random sampling and stratified sampling often result in clusters and gaps in a data set.  
Alternatively, Sobol sequence sampling \cite{saltelli_variance_2010}, a quasi-random sampling technique, offers a solution. Unlike other pseudo-random algorithms, Sobol sequence sampling avoids clustering and gaps even in smaller data sets \cite{renardy_sobol_2021}. It aims to generate multiple parameters uniformly distributed across a multi-dimensional parameter space.  
Thus, in the current work, Sobol sequence sampling generates a comprehensive design space. 10,000 data samples are generated with different combinations of the static features given in Table \ref{tab:materialparameters}. The resulting design space enables the exploration of various material and micro-structural configurations, providing valuable insights into the behavior of woven composite materials.

\begin{figure}[h!]
    \centering
    \captionsetup{justification=centering}
    \includegraphics[width=7cm]{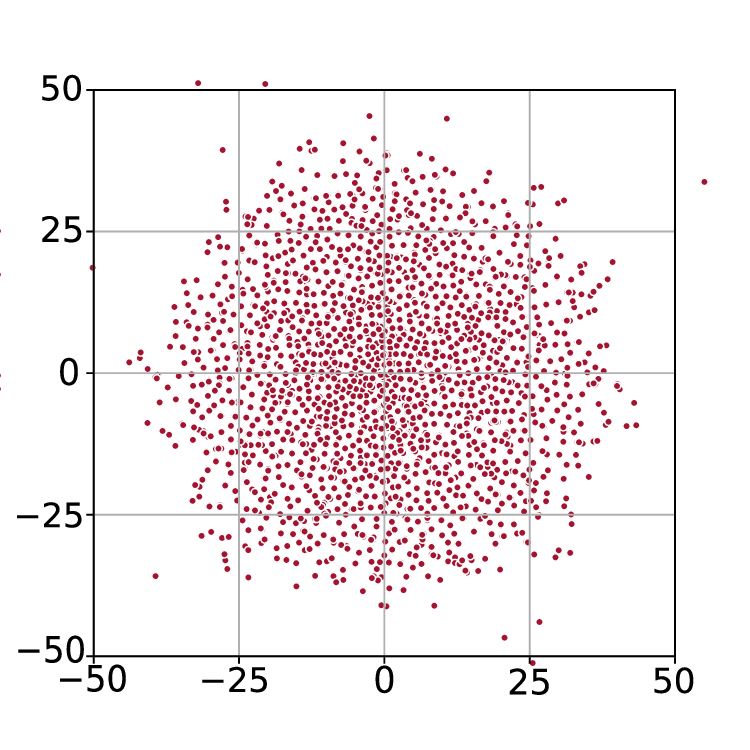}
    \vspace{-0.3cm}
    \caption{t-SNE distribution of 9-dimensional static feature space described in Table \ref{tab:materialparameters}. Different micro-structural configurations are described by three fiber elasticity, three matrix elasticity, and three matrix plasticity feature parameters.}
    \label{fig:t-SNE}
\end{figure}

For visualizing high-dimensional data sets, t-SNE (t-Distributed Stochastic Neighbor Embedding) is utilized \cite{vandermaaten08a, hinton2002stochastic}. It minimizes the discrepancy between high-dimensional and low-dimensional distributions using the Kullback-Leibler (KL) divergence \cite{mehlig_machine_2021} as a cost function. KL divergence quantifies how one probability distribution (from a multi-dimensional space) diverges from the expected (into a 2D space) probability distribution.
The algorithm iteratively adjusts data point positions to minimize the cost function. Monitoring KL divergence during cost minimization indicates capturing data structure and relationships in the lower-dimensional domain successfully. t-SNE preserves local structures, patterns, and clusters. While it is primarily a visualization tool, it can effectively represent data distribution, as shown in Figure \ref{fig:t-SNE}. 
No pattern or gap is visible on the scattered 2D plot, indicating a good correlation between regular grids and random distributions in the static feature space.

\subsubsection{Loading path generator}
\label{loadingPathGenerator}
In order to sample representative strain paths, different approaches can be employed, \eg \cite{heidenreich2023transfer, friemann2023micromechanics}. In the source task, a random walk representing long-term trends is combined with noises representing local variations. The algorithm \cite{friemann2023micromechanics} utilizes a six-dimensional space for independent strain components. Components are sampled independently from a normal distribution to generate direction vectors, then normalized to unit vectors. 
The algorithm defines parameters: $N_T$ as the total number of steps (constant and equal to 2000), $n_1$ as the number of drift directions (the number of major changes in the loading direction selected from $\{1,2,5,10,20,25,50,100,200\}$), $n_2$ as the noise vector with elements selected randomly from $\lfloor 0,1)$, and $\gamma$ as the perturbation amplitude factor (chosen randomly from $\lfloor 0,1)$). 

\begin{figure}[h!]
    \centering
    \captionsetup{justification=centering}
    \includegraphics[width=10cm]{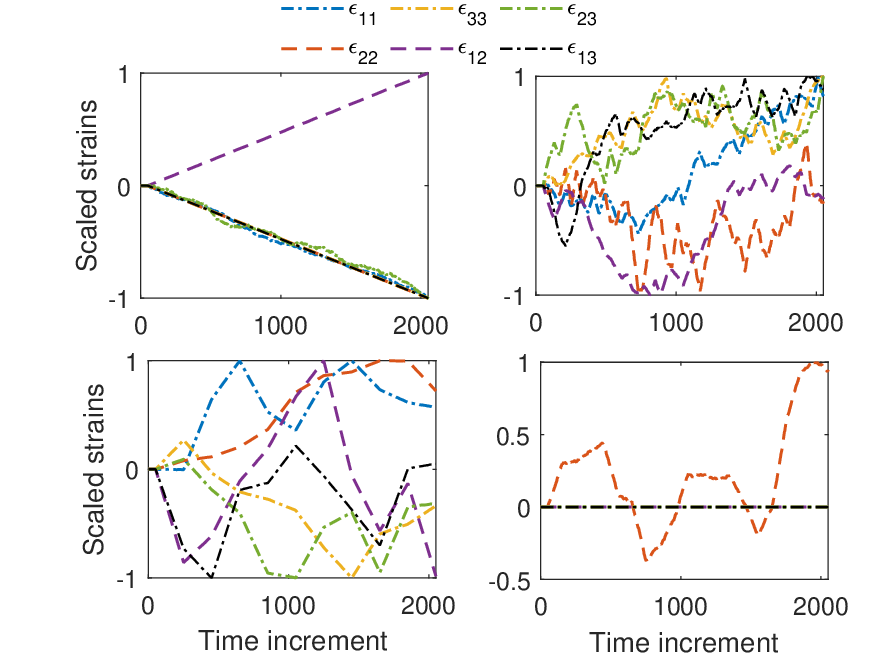}
    \vspace{-0.3cm}
    \caption{Four samples of input strain loading paths from random walk data set (scaled between [-1,1]). Each graph contains six components of the strain tensor applied on a randomly chosen material set.}
    \label{fig:randomStrain}
\end{figure}

Initially, $n_1$ drift directions are chosen randomly, and each of these directions is iteratively repeated $N/n_1$ times to form a vector comprising a total of $N_T$ elements. Subsequently, a noise vector, denoted as $n_2$, is generated with an equivalent number of elements as $N/n_1$ and is then scaled by a factor of $\gamma$. The vector of drift directions and the scaled noise vector are combined and scaled to a maximum of randomly selected $1\%$ to $5\%$ to complete one strain component of an input load sample. All components of a strain loading sample have the same number of $n_1$ but differ in other parameters. Figure \ref{fig:randomStrain} shows four samples of the generated loading paths. 
%

\begin{figure}[h!]
    \centering
    \captionsetup{justification=centering}
    \includegraphics[width=10cm]{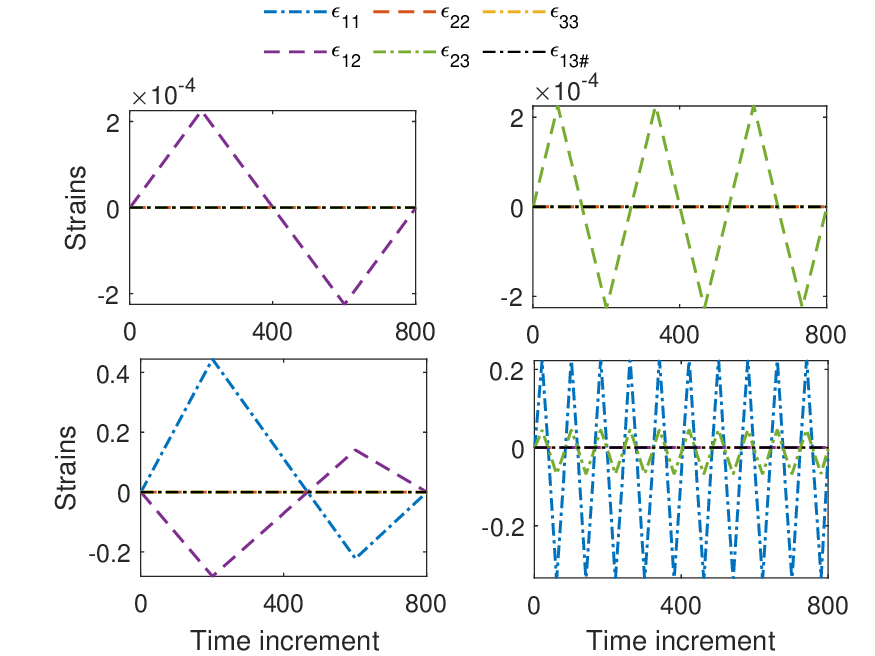}
    \vspace{-0.3cm}
    \caption{Four input strain loading paths samples from a smooth cyclic path data set. Each graph contains six components of the strain tensor applied on a randomly chosen material set. The smooth cyclic loading paths include (from up-left to down-right) pure in-plain shear $\epsilon_{12}$, out-of-plain shear $\epsilon_{23}$ and bi-axial $[\epsilon_{11},\epsilon_{12}]$ and $[\epsilon_{12},\epsilon_{23}]$ loading.}
    \label{fig:uniformStrain}
\end{figure}

Some loading samples have sparsity in their input features, such as the fourth case in Figure \ref{fig:randomStrain}, where the input strain tensor only includes the transverse strain component, while all other components are zero. Ince this rarely occurs, the trained network based on such a data set may need help to generalize the solution to cases with high feature sparsity, such as conventional cyclic loadings. Therefore, a second data set is generated based only on cyclic loads in shear and bi-axial load cases where plasticity is significant in woven composites. Figure \ref{fig:uniformStrain} shows four samples from the second data set. Indicating factors in the cyclic loads include the peak strain value, the strain ratio (ratio of shear strain to tensile strain in bi-axial scenarios), the load ratio (fraction of the maximum positive strain to the minimum negative strain value), and the number of cycles.


\section{Recurrent neural networks }
\label{RNN}
RNNs excel in handling data sequences of long lengths, making them suitable for language modeling, speech recognition \cite{chan2015listen}, machine translation \cite{cho2014learning, sutskever2014sequence}, and time-series prediction \cite{mehlig_machine_2021}.
RNNs effectively model sequential data due to their internal memory, capturing temporal dependencies. Feedback loops in the RNN architecture enable information flow between input and output, maintaining an internal state for predictions based on current and past inputs.

\subsection{Transfer learning rescuing RNNs from initialization and sparsity hurdles}
\label{transferLearning}

In the transfer learning paradigm, a model initially trained on a source task with abundant data serves as a foundational framework for a closely related target task. The primary objective is to leverage the pre-trained model's acquired knowledge of features, representations, and patterns to enhance the learning process on the target task \cite{yang2020transfer}. Furthermore, during the training phase for the source task, pre-trained models undergo regularization mechanisms, which mitigate overfitting risks. Transfer learning entails initializing the model with optimized weights specific to certain features, resulting in noteworthy acceleration of convergence and optimization improvements tailored for the target task. Transfer learning has been proven helpful for data fusion in elastic regime \cite{callaghan2023quantitative} and elasto-plastic behavior of short-fiber composites \cite{jung2022transfer}.

One specific strategy within transfer learning is fine-tuning \cite{heidenreich2023transfer}. This approach is employed when the source and target tasks are closely related, allowing for the adjustment of weights in a pre-trained model based on the data specific to the target task. In the context of this study, the source task involves predicting the six components of stress sequences derived from random loading simulations. Subsequently, during the original network training, the target task is formulated to predict stress components associated with cyclic loading paths. The RNN model employed for the source task integrates an a priori model, initially trained on extensive data sets, to adapt and enhance its performance on novel tasks. The weights and biases derived from the neural networks at the end of training serve as the initialization for the neural networks used in transfer learning.

\subsection{Recurrent learning with GRU and LSTM units}
\label{GRU_LSTM}
Gated Recurrent Units (GRUs) and Long Short-Term Memory (LSTM) networks are two types of recurrent neural networks (RNNs) that address the vanishing and exploding gradient problems encountered in traditional RNNs \cite{pascanu2013difficulty, lipton2015critical, chung2014empirical}. GRUs use two so-called gating mechanisms, an update gate, and a reset gate, to selectively update and reset their hidden state. This enables them to gather information across time steps and retain long-term dependencies \cite{cho2014properties}. LSTM units, on the other hand, employ three gating mechanisms: the input gate, the forget gate, and the output gate. These gate mechanisms control the flow of information, select information to retain or discard, and filter relevant information for output, respectively \cite{hochreiter1997long}. An interested reader is referred to \cite{matlabdoc} for a more comprehensive understanding of LSTM and GRU units. 

\begin{figure}[h!]
    \centering
    \captionsetup{justification=centering}
    \includegraphics[width=\linewidth]{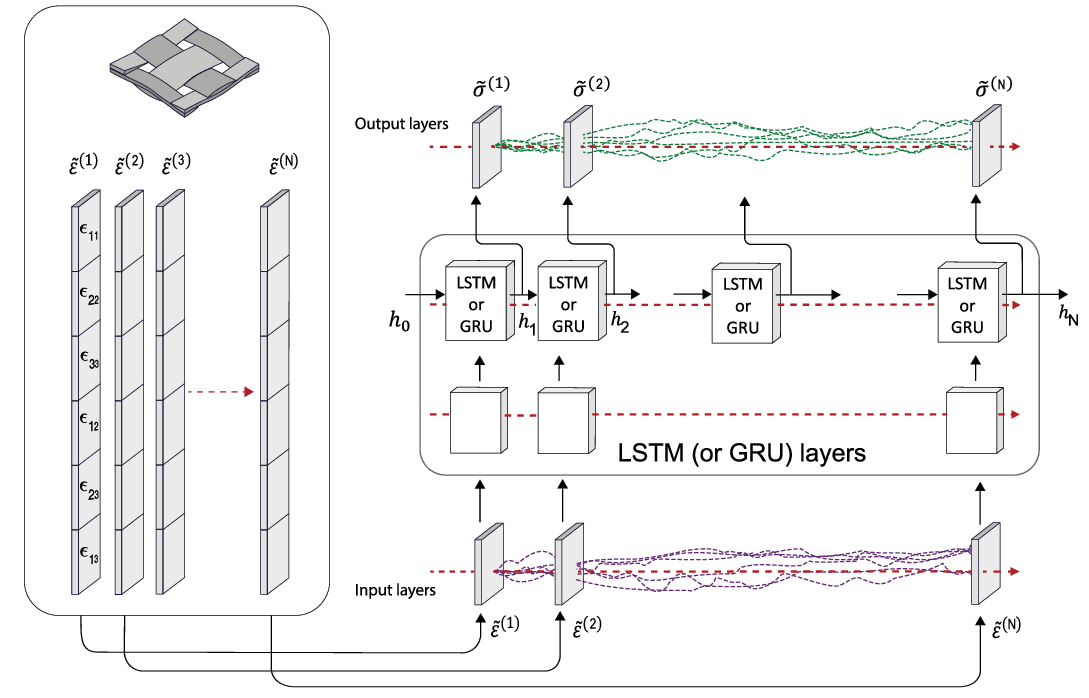}
    \vspace{-0.3cm}
    \caption{A schematic representation of the data flow for feeding the RNN model with one multiaxial strain history. As shown on the left side of the figure, a woven composite and strain loading paths have been randomly sampled from the data set. Each stripe represents a time step of the strain tensor ($\boldsymbol{\tilde{\epsilon}^{(n)}}$) containing six strain components ($\epsilon_{ij}$). Red dashed lines indicate time increments in a load path. Six random loads are presented in the input layer in purple. Afterwards, the data propagates through the LSTM (or GRU) network layers. The network outputs are in green. The internal variables flow can also be seen in one of the network layers. A hidden (and/or cell) state is indicated by $h_{t}$. }
    \label{fig:flowChart}
\end{figure}

Since GRUs have only two gates, they are computationally more efficient and easier to train with fewer parameters. However, they may not be as effective at capturing complex long-term dependencies as LSTMs. LSTMs with three gating mechanisms can model more complex relationships over longer sequences, but they have a higher computational cost and require more training data to prevent overfitting \cite{geron2022hands}. The choice between GRUs and LSTMs depends on the specific task at hand, the availability of computational resources, and the desired trade-off between model complexity and performance. To better understand how data flows through an RNN network containing two layers of recurrent units, Figure \ref{fig:flowChart} is provided. 
%
Learning rate, minibatch size, regularization strength, dropout rate, and network architecture are crucial parameters during RNN training. This is especially critical before the transfer learning process on the target task. Regularization (also called weight decay) \cite{bishop2006pattern} adds a penalty term to the loss function, promoting smaller weights in the network. This prevents the model from becoming overly sensitive to training data and enhances its generalization ability to unseen data. Additionally, the dropout layer \cite{JMLR:v15:srivastava14a} randomly sets a fraction of input units to zero during training, preventing co-adaptation of neurons and encouraging the network to learn more robust and independent representations. Combining $L_2$ regularization with dropout mitigates overfitting, reduces model complexity, and fosters more generalizable representations \cite{murphy2012machine}. Gradient clipping, setting the absolute value to one or using the maximum of normalized features, helps eliminate exploding gradients \cite{pascanu2013difficulty}. 

\subsection{Feature scaling}
\label{featureScaling}
Differences in input variable magnitudes can bias neural networks and hinder network learning. Variables with larger values dominate and overshadow smaller ones, leading to unstable weight updates and sub-optimal performance. Standardization equalizes input feature scales, ensuring smoother convergence in the network training and preventing disproportionate variable influence. It also allows direct magnitude comparison and enhances interpretability, thereby promoting stable training, faster convergence, and improved neural network performance \cite{mehlig_machine_2021}.

The standardization process involves normalizing sequential strain components to an absolute maximum value of one. Non-sequential material properties undergo min-max scaling to have values between zero and one. This enables neural networks to capture the underlying patterns better, resulting in improved performance and reliable predictions.

The network input consists of sequential data with $15$  different features, concatenating  $6$ sequential strain tensor components and $9$ static material properties. In each instance, the non-sequential features (the material properties) are repeated throughout the entire sequence of  $N_T = 2000$ pseudo-time increments. 


\subsection{Metrics beyond loss}
\label{metrics} 
Network performance relies not only on architecture but also on hyperparameters like learning rate, regularization strength, batch size, etc. The choice of hyperparameters can significantly impact the validation loss, potentially overshadowing the effect of selecting the optimal network architecture based solely on validation loss. 

This study considers additional evaluation metrics to find an optimal network configuration. The von Misses stress definition in Equation \ref{eq:vonMisesSigma} is adopted to represent all output stress components.
Using the von Mises effective stress, a single sequential vector is computed. 

Three statistical measures \cite{willmott2005advantages} are used to assess and compare the predictive performance of multiple neural network configurations. In the following, $e_i^{(t)}$ represents an individual model-prediction (for one sample) error at time step \(t\), defined as
\begin{equation}
\label{eq:Error}
    e_i^{(t)} = \hat\sigma_{i_{vM}}^{(t)}-{\sigma}_{i_{vM}}^{(t)},
\end{equation}
where $\hat{\sigma}^{(t)}_{vM}$ is the predicted equivalent stress at time step $t$ and $\sigma^{(t)}_{vM}$ is the true equivalent stress at $t$.
\\
\\
\noindent \textit{\textbf{Mean Absolute Error}}
\\
The mean Absolute Error (MAE) measures the average magnitude of errors over the data sequence length ($N_T$) between the predicted and desired values. MAE is calculated by taking the average of the absolute differences between each predicted and desired value, normalized by the number of tested samples from the unseen data set ($M$):
\begin{equation}
\label{eq:MAE}
    \text{{MAE}} = \frac{1}{MN_T}\sum_{m=1}^{M}\sum_{t=1}^{N_T} \left|e_i^{(t)}\right|.
\end{equation}
MAE provides an indication of the average size of errors produced by the model. There is no consideration for the direction of errors (overestimation or underestimation), and all errors are given equal weight. A lower MAE indicates a better performance. 
\\
\\
\noindent \textit{\textbf{Root Mean Square Error}}
\\
Another commonly used measure of prediction error is the Root Mean Square Error (RMSE). It is obtained by taking the square root of the average of the squared differences between the predicted values and the desired values as
\begin{equation}
\label{eq:RMSE}
    \text{{RMSE}} = \sqrt{\frac{1}{MN_T}\sum_{m=1}^{M}\sum_{t=1}^{N_T} (e_i^{(t)})^2}.
\end{equation}
Due to the squaring operation, larger errors are penalized more heavily. RMSE gives a measure of the overall deviation or dispersion of errors. Similarly to MAE, a lower RMSE indicates a better performance. Similarly to MAE, RMSE does not consider whether the model overestimates or underestimates a quantity at a given time instant.
\\
\\
\noindent \textit{\textbf{Mean Bias Error}}
\\
The bias or Mean Bias Error (MBE) refers to the systematic deviation between model predictions and the desired values in a data set. Bias can be positive or negative, indicating whether the predictions consistently overestimate or underestimate desired values. The scale and units of the predicted and desired values typically determine the bias range. MBE is computed by
\begin{equation}
\label{eq:MBE}
    \text{{MBE}} = \frac{1}{MN_T}\sum_{m=1}^{M}\sum_{t=1}^{N_T} e_i^{(t)}.
\end{equation}

\section{Results and discussion}
\label{Results and discussion}

For data consistency, only completed simulations are used\footnote[1]{Some of the simulations with the highest number of loading drifts (200) are not converged in the Digimat-MF solver.}. 
The output signals are components of the stress tensor. The mean squared error loss function is computed at the regression layer and is given by
\begin{equation}
    \label{eq:matlabcost}
    L = \frac{1}{N_{\text{batch}}}\sum_{k=1}^{N_{\text{batch}}}L_k,\quad \text{with}\quad L_k = \frac{1}{2N_T}\sum_{i=1}^{F} \sum_{j=1}^{N_T}  (\hat{\sigma}^{k}_{ij}-\sigma^{k}_{ij})^2,
\end{equation}
where $\hat{\sigma}^{k}_{ij}$ and $\sigma^{k}_{ij}$ are the $k^{th}$ stress predictions and target values in a training batch, respectively, $F$ is the number of output features, $N_T$ is the data sequence length, and $N_{batch}$ is the batch size. The computational programming language MATLAB \cite{matlabdoc} is employed for implementing recurrent neural networks. Training for different network configurations is done on the Vera hardware \cite{C3SEWebpage} at Chalmers University of Technology. 

Two data sets are generated, including 28,000 data samples for the source task on random walk loading paths and 10,000 samples for the target task on smooth cyclic loading paths. Each data set is randomly split into training (80\%), validation (10\%), and test (10\%) sets. For each task, the training data set is iteratively passed through the neural network for multiple epochs, with shuffled data at each epoch. The validation set helps for tuning hyperparameters, such as batch size, learning rate, and regularization parameters, to achieve an optimal model performance. The test set serves as unseen data for final performance evaluation after the network training and validation.

Section \ref{predictionOnRandomTestSet} presents results related to the source task. The purpose of Section \ref{hyperparameters} is to emphasize the importance of tuning the network architecture and hyperparameters. Once the network's generalization ability is confirmed for a random load path, the transfer learning method is used in Section \ref{predictionCyclicLoads} for training a network to predict smooth cyclic loads (target task).

\subsection{Prediction on the random strains test samples}
\label{predictionOnRandomTestSet}
\begin{figure}[h!]
    \centering
    \captionsetup{justification=centering}
    \includegraphics[width=8cm]{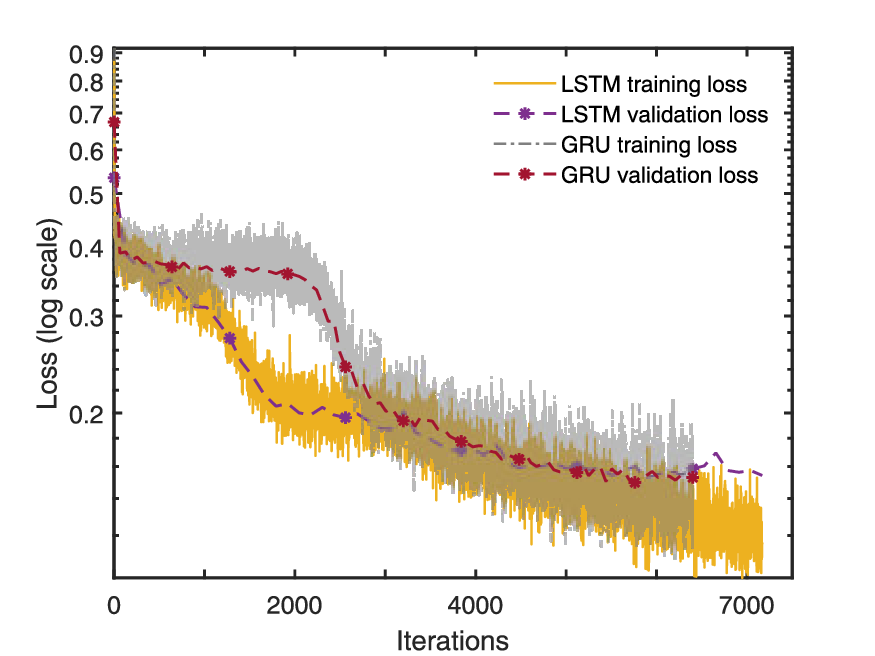}
    \vspace{-0.3cm}
    \caption{Two examples of the training and validation loss evolutions on the random strain data set. Each network consists of three layers with 512 units of LSTM or GRU plus a $50\%$ dropout after the first layer. In both cases, the learning rate is 0.001, the $L_2$ regularization is set to 0.001, and the minibatch size is 128. Yellow and gray lines indicate the training loss calculated at each iteration. The dashed lines in red and purple indicate the loss calculated at the end of each epoch for the validation set.}
    \label{fig:LossRandom}
\end{figure}

A piece-wise learning rate decay strategy is employed, reducing the learning rate by $10\%$ every ten epochs to enhance convergence. The training process uses ADAM optimizer \cite{kingma2014adam}. In order to prevent overfitting, early stopping is used instead of a fixed number of epochs. Training stops when the model's performance on the validation set plateaus, while the loss on the training set continues to decrease. Figure \ref{fig:LossRandom} illustrates the loss evolution for LSTM and GRU networks in accordance with Equation \ref{eq:matlabcost}. The plateau region on the validation set indicates convergence in the training of both cases. While the GRU network struggles more to reach the minimum loss value at the beginning, the minimum loss values are almost equal.

\begin{figure}[h!]
    \centering
    \captionsetup{justification=centering}
    \begin{subfigure}[b]{0.6\textwidth}
        \centering
        \includegraphics[width=\linewidth]{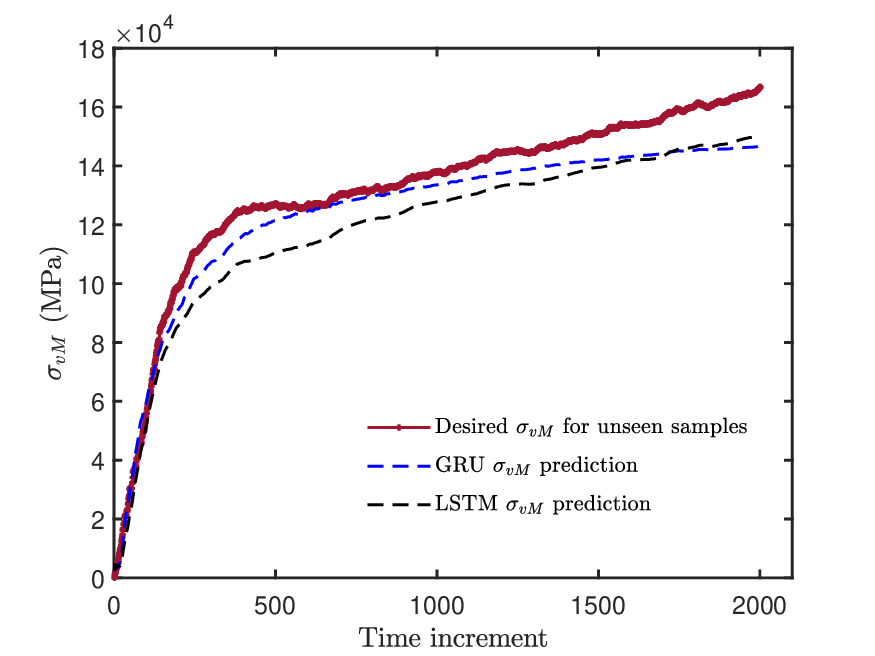}
        \caption{}
        \label{fig:vonMisesRandomPrediction1}
    \end{subfigure}
    \vspace{0.5cm} 
    \begin{subfigure}[b]{0.6\textwidth}
        \centering
        \includegraphics[width=\linewidth]{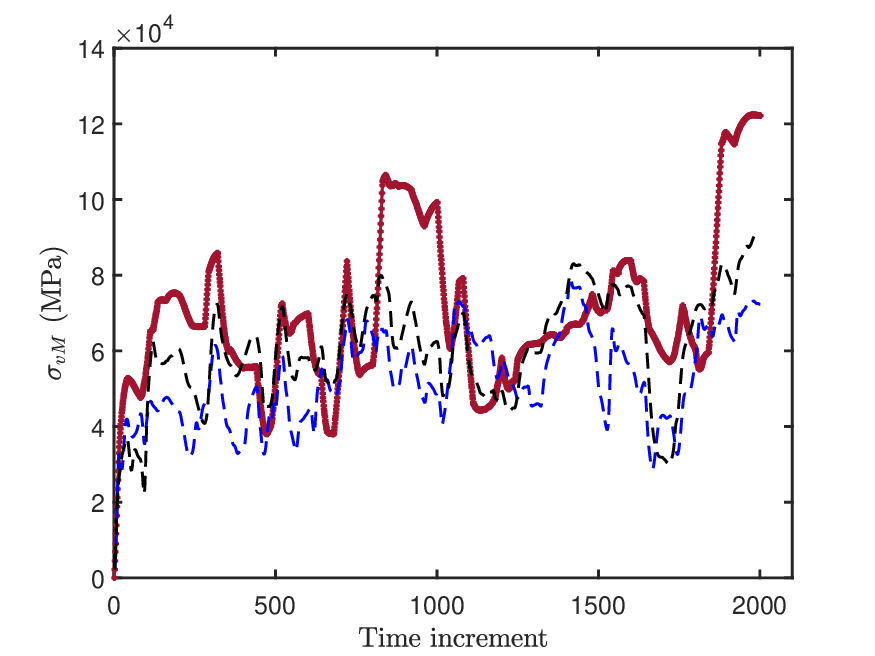}
        \caption{}
        \label{fig:vonMisesRandomPrediction2}
    \end{subfigure}
    \caption{Candidate LSTM (dashed black lines) and GRU (dashed blue lines) predictions on two samples with (a) $n_1 = 1$ and (b) $25$ drift directions from the random walk paths of the unseen data set. The von Mises stresses are plotted on the real scale for two distinct sets of micro-structural constituents material properties.} 
    \label{fig:vonMisesRandom}
\end{figure}

The three metrics, defined in Section \ref{metrics}, are computed based on the predictions made for each sample in the test set. 
The mean value over pseudo-time steps for three error metrics is obtained and presented in \href{https://chalmers-my.sharepoint.com/:b:/r/personal/ghane_chalmers_se/Documents/supplementary_materials_RNN_MF.pdf?csf=1&web=1&e=iNy6W7}{Tables 1} and \href{https://chalmers-my.sharepoint.com/:b:/r/personal/ghane_chalmers_se/Documents/supplementary_materials_RNN_MF.pdf?csf=1&web=1&e=iNy6W7}{2} in supplementary materials for different network architectures and the number of training parameters. Various configurations of GRU and LSTM networks are presented in these tables in order to determine the best architecture. The network configuration is defined by the number of GRU (or LSTM) units and the specification of the subsequent dropout layer. For example, the "3GRU(128) dp(40\%)" network consists of three GRU layers, each containing 128 units, followed by a 40\% dropout layer. For detailed error evaluations on GRU and LSTM networks, the reader is referred to the supplementary material of this study.
Our objective is to determine the optimal architecture first; afterward, various hyperparameters are discussed using LSTM or GRU units in Section \ref{hyperparameters}. Based on von Mises stress, Figure \ref{fig:vonMisesRandom} compares two best-performing LSTM and GRU networks in predicting two random walk loads from the unseen data set.

Although the smallest network "GRU(32)," had the lowest validation loss, it can not generalize well due to higher MAE and RMSE error ranges when testing unseen test data. "3LSTM(512) dp($50\%$)" displayed the lowest MAE and RMSE, indicating better prediction and generalization on unseen data than other models. However, note that the MBE of "3LSTM(512) dp($50\%$)" is not closest to zero, and its negative value suggests a systematic underestimation of equivalent stress.

\begin{figure}[h!]
    \centering
    \captionsetup{justification=centering}
    \begin{subfigure}[b]{0.6\textwidth}
        \centering
        \includegraphics[width=\linewidth]{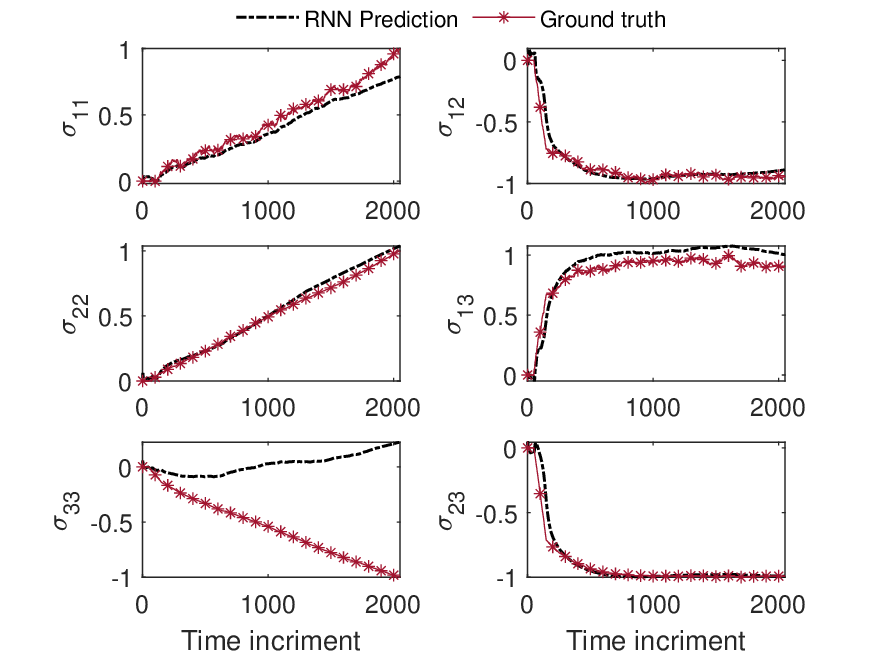}
        \caption{}
        \label{fig:signal1}
    \end{subfigure}
    \vspace{0.5cm} 
    \begin{subfigure}[b]{0.6\textwidth}
        \centering
        \includegraphics[width=\linewidth]{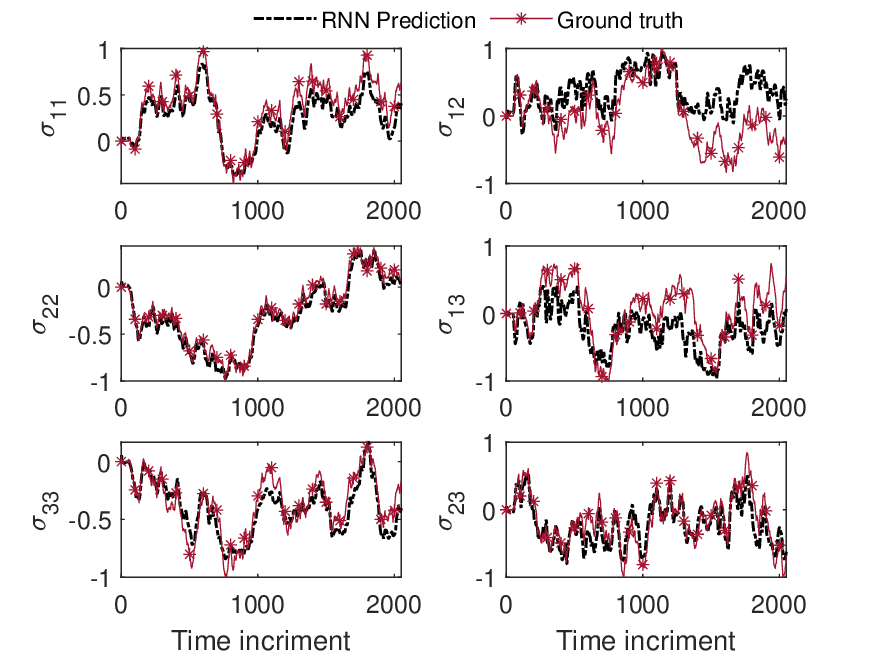}
        \caption{}
        \label{fig:signal2}
    \end{subfigure}
    \caption{Network (3LSTM(512) dp($50\%$)) predictions against micro-mechanical results for two samples from unseen random load cases. Stress values are scaled between $[-1,1]$ (a) Six components of stresses in a case with a rather uniform load and (b) a case with high randomness in loading. The solid red line is the mean-field simulation output (desired values), and the dashed black lines are the network predictions.} 
    \label{fig:unseenStress}
\end{figure}

Figure \ref{fig:unseenStress} shows predictions for stress components on two randomly selected test samples for the candidate network based on error metrics "3LSTM(512) dp($50\%$)". 
Figure \ref{fig:signal1} shows a strong correlation between the predictions and the desired stress values. However, stress prediction deviates for $\sigma_{33}$, where the desired value is close to zero throughout the loading increments. It is potentially related to the general underlying issue with neural networks, known as feature sparsity \cite{mehlig_machine_2021}, which needs further research to handle sequential regression tasks. A network might have difficulty predicting features that are close to zero in a sample when there is a high number of input features (15 in this case). 

In spite of a highly random loading path, in Figure \ref{fig:signal2}, the predicted values match well with the desired values from mean-field simulation. It can be seen that the predictions on normal components are better than those on shear components. This observation could be related to the higher level of non-linearity in shear components.
%

\subsection{Discussion on hyperparameters}
\label{hyperparameters}
Various networks have undergone grid search training to find the best hyperparameter combination. The tested hyperparameters have been examined as follows: Minibatch Size = [16, 32, 64, 128], $L_2$ Regularization = [0.001, 0.01, 0.], Dropout Rate = [0.2, 0.5, 0.8], Learning Rate = [0.0001, 0.001, 0.01]. The grid search method leads to training and evaluating more than 200 LSTM and GRU networks with different hyperparameter combinations.
For detailed error evaluations, please refer to \href{https://chalmers-my.sharepoint.com/:x:/r/personal/ghane_chalmers_se/Documents/NetworksEvaluationWithVonMisesStress.xlsx?d=w097efe717f8e473e88bfcb15528fde29&csf=1&web=1&e=4KOZcm}{Table 3} in the supplementary material of this study.

The candidate LSTM and GRU networks have three layers, each containing 512 units. As a result of the hyperparameters grid search, the LSTM network's optimum learning rate and minibatch size are 0.001 and 128, respectively, while the candidate GRU network optimum is 0.1 and 32. The regularization and the dropout rate are optimal in both networks at 0.001 and $50\%$, respectively.
While LSTM networks outperform GRU networks in minimizing RMSE and MAE, they come with the cost of, on average, 15 times longer training times. 

\subsection{Prediction on conventional load cycles}
\label{predictionCyclicLoads}
The initial task, referred to as the source task, involves predicting loading cases associated with random walk loading paths. Subsequently, the focus shifts to the target task, where the objective is to predict loading paths characterized by conventional cyclic patterns. 

\begin{figure}[h!]
    \centering
    \captionsetup{justification=centering}
    \includegraphics[width=8cm]{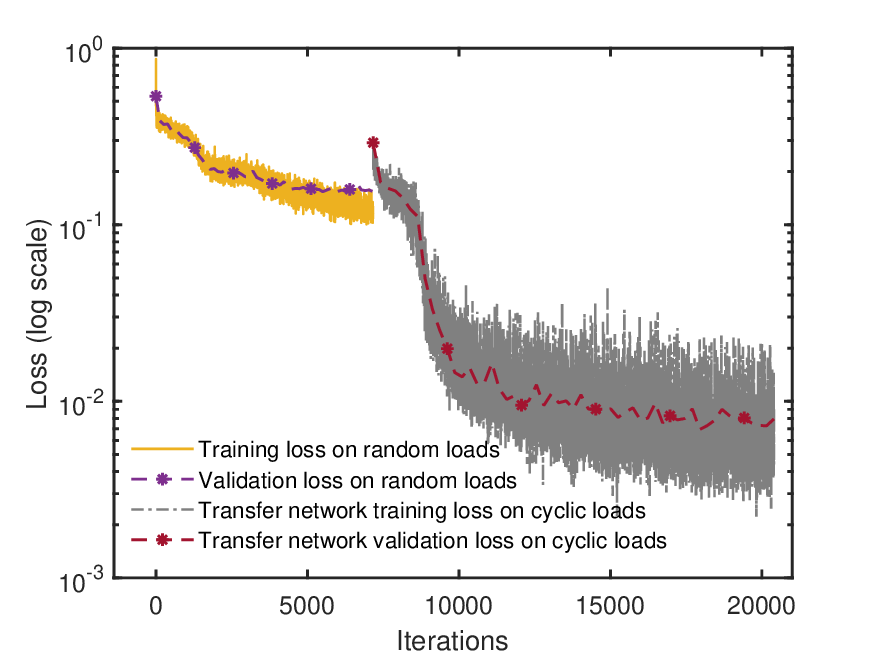}
    \vspace{-0.3cm}
    \caption{On the source task, the original LSTM network training and validation losses are presented in yellow and dashed purple, respectively. Transfer learning is then applied to the target task (cyclic loads), and the loss is computed (gray for training and red for validation).}
    \label{fig:LossTransfer}
\end{figure}

However, a notable challenge is associated with applying the trained network to the target task involving cyclic loads. The input features of cyclic loading samples consist of only one (for pure shear cases) or two (for bi-axial loading cases) active sequential features in addition to the static features (micro-structure). As a result, the cyclic loading features are sparse, and the ANN models have difficulty predicting stresses based on sparse samples. Therefore, the network undergoes fine-tuning with a data set specific to the target task. As can be seen in Figure \ref{fig:LossTransfer}, the evolution of loss illustrates the convergence of the transfer learning process following the initial training of the LSTM network. As a result of the first 7000 iterations, the network loss function converged on both the training and test sets of the random load paths data set. The loss function jumps at the beginning of the fine-tuning process when samples from cyclic loads are fed to the network. The network is then trained using cyclic loads samples through more iterations, and convergence occurs after 20,000 iterations.

\begin{figure}[h!]
    \centering
    \captionsetup{justification=centering}
    \begin{subfigure}[b]{0.6\textwidth}
        \centering
        \includegraphics[width=\linewidth]{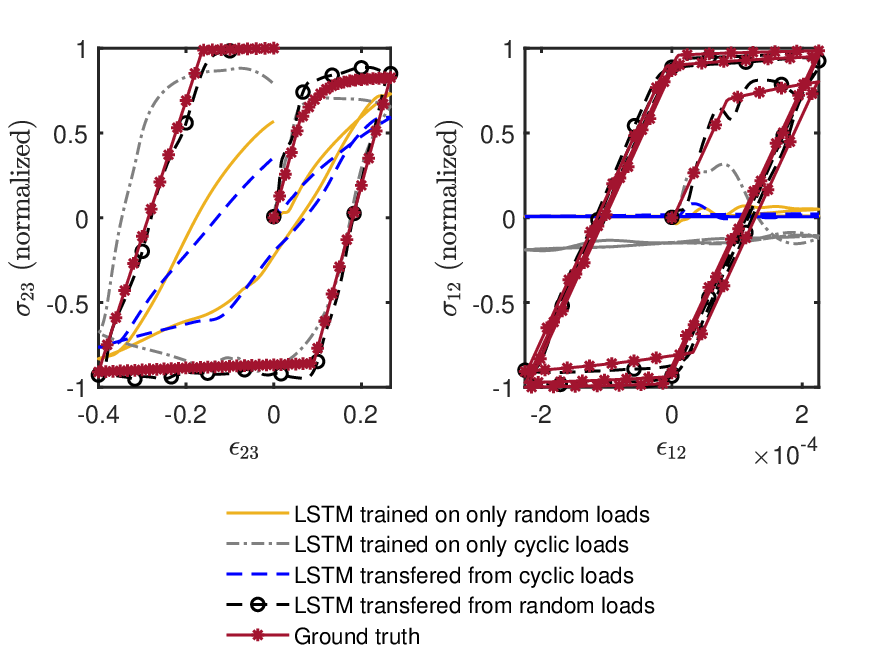}
        \caption{}
        \label{fig:cycle1Pred}
    \end{subfigure}
    \vspace{0.5cm} 
    \begin{subfigure}[b]{0.6\textwidth}
        \centering
        \includegraphics[width=\linewidth]{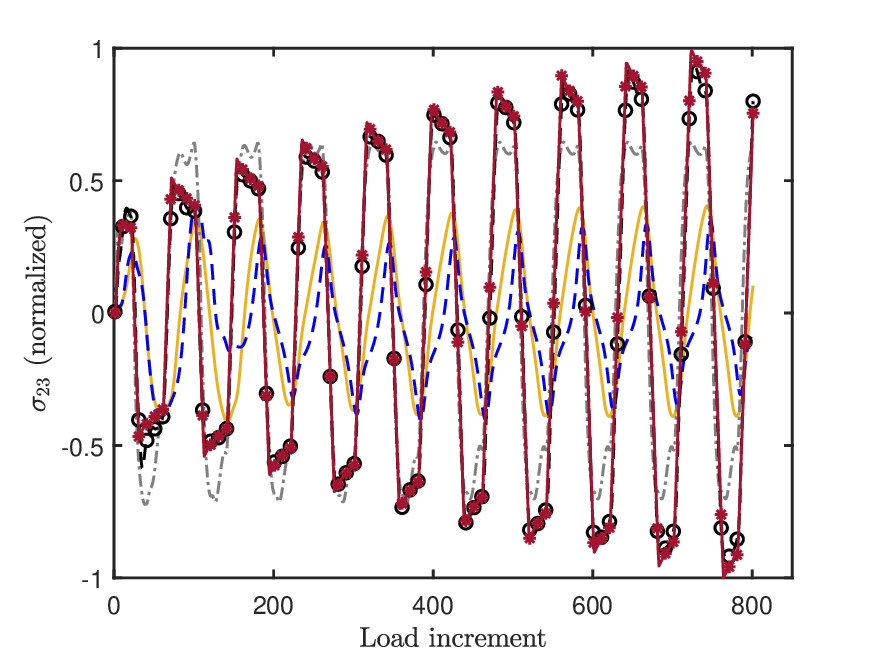}
        \caption{}
        \label{fig:cycles2Pred}
    \end{subfigure}
        \vspace{0.5cm} 
    \caption{LSTM network predictions on conventional cyclic loading compared with desired stress values (red solid lines with stars). Stress values are normalized between $[-1,1]$. Black dashed lines indicate the candidate transferred model predictions. The blue dashed lines indicate predictions of a model trained initially on cyclic loads and then fine-tuned by random loads. The yellow and grey lines show the predictions of models that have been trained only on random and cyclic loads, respectively. (a-left) one cycle pure out-of-plain shear, (a-right) three cycles pure in-plain shear. (b) ten cycles pure out-of-shear vs. load increments.} 
    \label{fig:cyclicPredictions}
\end{figure}

The original network is trained on random walk paths and then fine-tuned on cyclic loading. There are two possible alternative scenarios: (1) the original network is limited to being trained on cyclic loads, or (2) the original network is initially trained on cyclic loads and fine-tuned with random walk paths. Three instances of cyclic loads are shown in Figure \ref{fig:cyclicPredictions}, corresponding to a one-cycle test, a three-cycle test, and a ten-cycle test, respectively. 
It can be clearly seen that the original network trained based on the random walk data set (represented by the solid yellow line) can not predict the cyclic loads despite its performance on the random walk data set. The dashed grey line indicates that the data set for the cyclic load is insufficient and that most of the features are too sparse to train a network to predict unseen test targets. The dashed blue line illustrates an alternative approach that involves training the network on cyclic loads and fine-tuning it using random walks. According to the results, transfer learning, in this case, cannot perform satisfactorily.

The most promising results are obtained by training the model on random walk load paths (Section \ref{predictionOnRandomTestSet}) and then fine-tuning it for cyclic loads, as illustrated by the dashed black line. With the transferred network, stress values can be predicted from sparse feature samples through the entire loading increments not only for one-cycle tests (Figure \ref{fig:cycle1Pred}-left) but also for multiple cycles (Figure \ref{fig:cycle1Pred}-right and \ref{fig:cycles2Pred}). In some cases, the predictions deviate from the desired values at the beginning of loading, as shown in Figure \ref{fig:cycle1Pred}-right. This behavior can be attributed to the nature of RNNs, which forget about the first part of a sequence when asked to predict the whole sequence at once, and not for forecasting and predicting one step ahead \cite{geron2022hands}.

\section{Conclusions}
\label{Conclusions}

Woven composite laminates are used across a wide range of industries due to their cost-effective and automated manufacturing procedures. However, computationally efficient and industrially feasible models are still required to drive their use further.
This study has investigated the capabilities of Recurrent Neural Networks (RNNs) combined with the transfer learning strategy in supervised learning of the nonlinear behavior of woven composites, specifically when plasticity is present in the matrix phase. The present research has examined the predictive power of RNNs based on mean-field simulated data to be used as a fast and accurate surrogate model for mesoscale homogenization. 
The RNN model incorporates two types of input features: (1) static input features sampled from a random design space for elastic fiber and elasto-plastic matrices with varying volume fractions; (2) sequential input features including six-dimensional time histories of the meso-scale strain tensor. The RNN model aims to predict six components of meso-scale homogenized stresses as outputs. 
Two distinct task data sets are considered. In the first one, named the source task, a random walk strategy ensures a diverse and comprehensive exploration of input strain path trajectories. In the second one, named the target task, conventional cyclic strain loadings are considered. The second task was found to be more challenging due to the presence of multiple zero-vectors in the sequential input features, known as feature sparsity. 
We have systematically examined a variety of GRU and LSTM architectures, along with several hyperparameters, in order to identify the most suitable model for predicting homogenized stresses in unseen test samples derived from random loads (source task). Having been exposed to random loads, the network has been retrained using a transfer learning paradigm and has demonstrated satisfactory performance in predicting stress components under conventional cyclic load conditions (target task). 

In conclusion, this study provides evidence for transfer learning neural networks as an effective method for domain adaptation across diverse loading types and micro-structural constituent properties in material data sets.
Our findings set a precedent for future investigations into knowledge transfer across varying-quality datasets. The results provide a robust foundation for future studies to tailor models for full-field simulations and even limited experimental data. The insights gained from mean-field simulations can be used to develop models that are applicable in a wider range of settings and represent a significant opportunity for further advancement in this field.

\section*{Acknowledgment}
\noindent S.M. Mirkhalaf and E. Ghane gratefully acknowledge financial support from the Swedish Research Council (VR grant: 2019-04715) and the University of Gothenburg. M. Fagerström is thankful for the support through Vinnova's strategic innovation programme LIGHTer, in particular via the project LIGHTer Academy Phase 3 (grant no. 2020-04526). The computations were enabled by resources provided by Chalmers e-Commons. E. Ghane is also grateful to Joel Magnusson, Iuri Rocha and Hon Lam Cheung for their input and stimulating discussions, which enhanced this research.

\bibliographystyle{unsrt}
\bibliography{2.References}{}

\newpage


\end{document}